\documentclass[aps,prb,twocolumn,groupedaddress,showpacs]{revtex4}
\usepackage{epsfig}
\usepackage{graphicx}
\usepackage{color}

\begin{document}

\title{Isentropic Curves at Magnetic Phase Transitions}

\author{J.D. Cone, A. Zujev and R.T. Scalettar}

\affiliation{Physics Department, University of California, Davis,
California 95616, USA}

\begin{abstract}
Experiments on cold atom systems in which a lattice potential is ramped
up on a confined cloud have raised intriguing questions about how the
temperature varies along isentropic curves, and how these curves
intersect features in the phase diagram.  In this paper, we study the
isentropic curves of two models of magnetic phase transitions- the
classical Blume-Capel Model (BCM) and the Fermi Hubbard Model (FHM).
Both Mean Field Theory (MFT) and Monte Carlo (MC) methods are used.  The
isentropic curves of the BCM generally run parallel to the phase
boundary in the Ising regime of low vacancy density, but intersect the
phase boundary when the magnetic transition is mainly driven by a
proliferation of vacancies.
Adiabatic heating occurs in moving away from the phase boundary.
The isentropes of the half-filled FHM
have a relatively simple structure, running parallel to the temperature
axis in the paramagnetic phase, and then curving upwards as the
antiferromagnetic transition occurs.  However, in the doped case, where
two magnetic phase boundaries are crossed, the isentrope topology is
considerably more complex.
\end{abstract}

\pacs{
05.10.Ln, 71.10.Fd, 75.30.Kz,75.10.Jm,71.30.+h,64.60.Cn
}
%% Monte Carlo methods in statistical physics 05.10.Ln
%% Magnetic phase transitions, 75.30.Kz
%% Hubbard model magnetic ordering (quantized spin model), 75.10.Jm
%% Order-disorder transformations 64.60.Cn
%% Lattice fermion models (Hubbard model, etc.) 71.10.Fd
%% Metal-insulator transitions and other electronic transitions 71.30.+h

\maketitle

%%%%%%%%%%%%%%%%%%%%%%%%%%%%%%%%%%%%%%%%%%%%%%%%%%%%%%%%%%
%%%%%%%%%%%%%%%%%%%%%%%%%%%%%%%%%%%%%%%%%%%%%%%%%%%%%%%%%%
\section*{I.  Introduction}
%%%%%%%%%%%%%%%%%%%%%%%%%%%%%%%%%%%%%%%%%%%%%%%%%%%%%%%%%%
%%%%%%%%%%%%%%%%%%%%%%%%%%%%%%%%%%%%%%%%%%%%%%%%%%%%%%%%%%

In systems of interacting degrees of freedom, decreasing the thermal
fluctuations often leads to the formation of ordered states.  The
traditional, and natural, measure of these fluctuations is the
temperature $T$ itself, which then forms one axis of the associated
phase diagrams.  However entropy $S$ can also be used to quantify the
amount of disorder.  Indeed, a phase diagram using $S$ as an axis
naturally provides a somewhat different perspective on the topology of
the ordered and disordered regions- since the entropy changes more
rapidly where transitions occur, it magnifies these interesting portions
of the phase diagram.

Recent experiments\cite{jaksch98,greiner02,bloch04,lewenstein07}
on trapped ultracold atoms in optical
lattices have provided a further motivation for employing the entropy as
one of the variables in describing phase diagrams.\cite{optlatS,optlatS2}  In
these systems, the temperature and entropy of the atomic cloud are known
prior to the adiabatic ramp-up of the optical lattice, but
the precise change in temperature during this process is uncertain.
Thus the determination of the entropy values at which various phenomena
occur, like local moment formation, magnetic ordering, and so forth,
is important, supplementing the more typical discussion of the
temperatures at which these phenomena occur.  Furthermore, since
attaining low temperatures is crucial for the emulation of many-body
ordering effects seen in solid state systems, a central question is
whether $T$ rises or falls (adiabatic heating or cooling) as the optical
lattice is established.

This work examines the isentropic curves of two models of magnetic
phase transitions.  The two-dimensional Blume-Capel Model
(BCM)\cite{blume66,capel66} is studied first, and its thermodynamics are
computed both in Mean Field Theory (MFT) and with Monte Carlo (MC)
methods.  An itinerant (quantum) Hamiltonian, the Fermion Hubbard Model
(FHM) is next examined within MFT.  The isentropes of the FHM are
compared with recent Quantum Monte Carlo (QMC)\cite{paiva10} results at
half-filling, a density at which QMC can be performed to low
temperatures.  The isentropes are also computed when the system is doped
away from half-filling, a parameter regime where phases with long range
order are inaccessible to QMC.
%\color{blue}
The BCM and FHM form an interesting pair of models to compare, since
both contain two energy scales, one which controls the density
and the other which tunes the strength of spin-spin interactions.
%\color{black}

The paper is organized as follows:  Sec.~II presents both the BCM and
the FHM, and the computational methods used.  Sec.~III then details the
isentropes of the BCM, while Sec.~IV focused on the FHM.  Sec.~V
summarizes and further discusses the results.

%%%%%%%%%%%%%%%%%%%%%%%%%%%%%%%%%%%%%%%%%%%%%%%%%%%%%%%%%%
%%%%%%%%%%%%%%%%%%%%%%%%%%%%%%%%%%%%%%%%%%%%%%%%%%%%%%%%%%
\section*{II.  Models and Computational Approach}
%%%%%%%%%%%%%%%%%%%%%%%%%%%%%%%%%%%%%%%%%%%%%%%%%%%%%%%%%%
%%%%%%%%%%%%%%%%%%%%%%%%%%%%%%%%%%%%%%%%%%%%%%%%%%%%%%%%%%

%% The BCM will be studied both in MFT and with classical
%% Monte Carlo.  Our solution of the HH will be purely MF, but we will
%% compare results at half-filling with recent Quantum Monte Carlo (QMC)
%% simulations \cite{paiva10}.

%%%%%%%%%%%%%%%%%%%%%%%%%%%%%%%%%%%%%%%%%%%%%%%%%%%%%%%%%%
\subsection{Blume Capel Model}
%%%%%%%%%%%%%%%%%%%%%%%%%%%%%%%%%%%%%%%%%%%%%%%%%%%%%%%%%%

The Blume-Capel Model\cite{blume66,capel66} is,
\begin{eqnarray}
E = -J \sum_{\langle {\bf ij} \rangle} S_{\bf i} S_{\bf j}
+ D \sum_{\bf i} S_{\bf i}^2 \,\, .
\label{eq:BCM}
\end{eqnarray}
The spins $S_{\bf i}$ can take three values $S_{\bf i}=0,\pm1$, where
the value $S_{\bf i}=0$ can be regarded as a `vacancy'.  The first term
in the energy represents a ferromagnetic coupling $J$ between
near-neighbor spins.  The second term provides an 'impurity' chemical potential $D$
for the vacancies.  When $D$ is large and negative, vacancies are
suppressed, and the BCM maps onto the Ising model.  Here we will
consider a square lattice geometry.

The presence of a three component spin gives rise to
the possibility of first order transitions and tricritical
behavior, as first emphasized in [\onlinecite{griffiths73}].
Part of the original motivation for the
BCM was to provide a description of the tricritical phenomena
induced by $^3$He vacancies
in superfluid $^4$He.\cite{graf67,goellner71,ahlers74}
Since its introduction, the BCM has been extensively
studied,\cite{berker76,burkhardt77,saul74,jain80,wang87,hoston91,beale86,rachadi03}
both in the form given in Eq.~\ref{eq:BCM},
and in several variants which include additional
terms in the energy\cite{blume71,mukamel74}
and generalizations to vector spins
which more correctly capture the
continuous symmetry of the superfluid
order parameter.\cite{cardy79,berker79,chamati07,dillon10}

The solution of the BCM within MFT is straightforward.
The results for the free energy and entropy (per
site) are,
\begin{eqnarray}
f &=& \left(\frac{Jzm^2}{2}\right)
 - \frac{1}{\beta} \ln\ [ 2 \cosh (\beta J z m) + e^{\beta D}] + D
\label{eq:BCMsandf}
\\
s &=& -\beta [ J z m^2 + (1-\rho ) D]
+\ln [2 \cosh (\beta J z m) + e^{\beta D}]
\nonumber
\end{eqnarray}
where z is the coordination number ($z=4$ for square lattice).
The magnetization $m$ and
density of spin states $\rho$ are given by the self-consistency equations:
\begin{eqnarray}
m &=& \frac{2\sinh(\beta Jmz)}{e^{\beta D} + 2\cosh (\beta Jmz)}
\nonumber \\
\rho &=& \frac{2\cosh (\beta Jmz)}{e^{\beta D} + 2\cosh (\beta Jmz)}
\label{eq:selfconsist}
\end{eqnarray}
Here we set $J=1$ as a unit of energy, and solve
Eqs.~\ref{eq:BCMsandf},\ref{eq:selfconsist}
to obtain the
free energy and entropy for any values of independent variables D and $\beta$ .

We compared two different MC approaches for computing entropy in the BCM:
thermodynamic integration and the Wang-Landau (WL)
algorithm. With thermodynamic
integration, the entropy is computed by making multiple MC runs at
different, fixed inverse temperatures $\beta$ and integrating,
\begin{eqnarray}
s(\beta) = s(\infty) + \beta e - \int_\beta^\infty e(\beta) d\beta  \,\, ,
\label{eq:thermint}
\end{eqnarray}
where $s$ and $e$ are the entropy and energy per site, respectively.
Equation \ref{eq:thermint} is obtained by applying integration by
parts\cite{binder81,tremblay07,paiva10}
to the standard relation of entropy to the specific heat: $S(T) =
\int_0^T C(T')/T' \, dT'$.  For the BCM, $s(\infty)= \ln 3$, reflecting
the three possible choices of spin.  In most instances, this approach
produces better results, since it does not rely on the determination of
specific heat $C(T)$, which is noisier than the energy $e(T)$.

The WL algorithm is a flat-histogram
MC method for calculating the density of states
$g(E)$. In this method, a random
walk is performed in the energy space of the BCM, sampling $E$
%% with a probability proportional to $\frac{1}{g(E)}$ and adjusting
with a probability proportional to $1/g(E)$ and adjusting
the distribution of $g(E)$ until each energy (E) value has close
to the same probability. Ultimately, this process produces a flat
histogram of occurrence for all energy states in the random walk.
Since the density of energy
states that results is independent of temperature, we can compute
the partition function $Z= \sum_{E} g(E)e^{-\beta E}$ for any
temperature. Consequently, we can determine the value for any thermodynamic variable of
interest- in our case, the free energy and entropy- without performing
multiple MC simulations at different $\beta$.

Figures and analysis reported for BCM will be those from WL results
on 16x16 lattices. The values obtained for the entropy with the two
methods, however, were found to be equivalent to within a fraction of a percent.

%%%%%%%%%%%%%%%%%%%%%%%%%%%%%%%%%%%%%%%%%%%%%%%%%%%%%%%%%%
\subsection{Hubbard Hamiltonian}
%%%%%%%%%%%%%%%%%%%%%%%%%%%%%%%%%%%%%%%%%%%%%%%%%%%%%%%%%%

The Fermion Hubbard Model,\cite{hubbard63}
\begin{eqnarray}
H = &-&t \sum_{\langle {\bf ij}\rangle \,\sigma}
( c_{{\bf i}\sigma}^{\dagger} c_{{\bf j}\sigma}^{\vphantom{dagger}}
+ c_{{\bf j}\sigma}^{\dagger} c_{{\bf i}\sigma}^{\vphantom{dagger}} )
\nonumber \\
&+&U \sum_{{\bf i}} n_{{\bf i}\uparrow} n_{{\bf i}\downarrow}
-\mu \sum_{{\bf i}\sigma} n_{{\bf i}\sigma}  \,\, ,
\end{eqnarray}
describes the magnetism of itinerant electrons,
in contrast to the static spins of the Blume-Capel model.
Here
$c_{{\bf i}\sigma}^{\dagger}(c_{{\bf i}\sigma}^{\vphantom{dagger}})$
are creation(destruction) operators for fermions of spin $\sigma$
on lattice site ${\bf i}$,
and $n_{{\bf i}\sigma}$ are the associated number operators.
$t=1$ sets the kinetic energy scale for the hopping of fermions
between near neighbor sites of a square lattice.
$U$ is an on-site energy cost for double occupancy, and the chemical
potential $\mu$ controls the filling.

One interesting property of the square lattice, near-neighbor
hopping is that the associated dispersion relation
$\epsilon(k_x,k_y) = -2t [ {\rm cos}k_x + {\rm cos} k_y ]$
has a (logarithmic) divergent density of states
at half-filling.  As a consequence, the Stoner criterion
suggests that an arbitrarily small interaction $U$ will induce
a magnetic instability.
This is reflected in the phase diagrams shown in
Figs.~\ref{fig:Hirsch1} and \ref{fig:isenn1.0}.

The FHM has been widely used to study strong correlation effects in
solids, from magnetism to metal-(Mott) insulator transitions, and high
temperature superconductivity.\cite{hubbard63,rasetti91,montorsi92,fazekas99}
Recently, the FHM and its bosonic counterpart have attracted considerable
interest for describing the behavior of cold atoms trapped in an optical
lattice produced by interfering laser beams.\cite{jaksch98} Compared with
traditional condensed matter experiments, these optical lattice systems
are thought to more precisely mimic the FHM, while allowing tunable
 control over parameters like $U$ and $t$.
As mentioned in the introduction, this provides a central motivation for
this paper.

Our MFT approach is the usual one in which each term of
the interaction is decoupled:
$U n_{{\bf i}\uparrow} n_{{\bf i}\downarrow} \rightarrow
 U n_{{\bf i}\uparrow} \langle n_{{\bf i}\downarrow} \rangle
+U \langle n_{{\bf i}\uparrow} \rangle n_{{\bf i}\downarrow}
-U \langle n_{{\bf i}\uparrow} \rangle
   \langle n_{{\bf i}\downarrow} \rangle$.
The resulting quadratic Hamiltonians $H_\sigma$ are
diagonalized, and the expectation values
$\langle n_{{\bf i}\sigma} \rangle $ recomputed.
The process is iterated to self-consistency, a process which
minimizes the free energy $F$.
To be somewhat more precise, the MFT calculation is
actually performed in momentum space by making particular, simple,
paramagnetic (P), ferromagnetic (F), and antiferromagnetic (AF)
{\it ansatz} for the real space expectation values
$\langle n_{{\bf i}\sigma} \rangle$.
Such a choice does not allow for more complex spatial
patterns of charge and spin such as are present in
striped phases.\cite{zaanen89}
%% Because $U$ only enters in the combination
%% $U \langle n_{{\bf i}\sigma} \rangle $,
%% the free energy and entropy $S$ will be independent of $U$ in
%% in the paramagnetic phase if the density is held fixed.

Our results for the isentropic curves of the FHM
will be obtained exclusively within MFT.  It is
possible to obtain these curves using more exact approaches
like QMC,\cite{paiva10} but only in parameter regimes like half-filling
where there is no sign problem.%\cite{loh90}
Previous Dynamical Mean Field Theory work has also reported
data for the isentropes on a cubic lattice.\cite{werner05,optlatS2}

%%%%%%%%%%%%%%%%%%%%%%%%%%%%%%%%%%%%%%%%%%%%%%%%%%%%%%%%%%
%%%%%%%%%%%%%%%%%%%%%%%%%%%%%%%%%%%%%%%%%%%%%%%%%%%%%%%%%%
\section*{III.   Isentropic Curves of the Blume-Capel Model}
%%%%%%%%%%%%%%%%%%%%%%%%%%%%%%%%%%%%%%%%%%%%%%%%%%%%%%%%%%
%%%%%%%%%%%%%%%%%%%%%%%%%%%%%%%%%%%%%%%%%%%%%%%%%%%%%%%%%%

%%%%%%%%%%%%%%%%%%%%%%%%%%%%%%%%%%%%%%%%%%%%%%%%%%%%%%%%%%
\subsection*{Single Site Limit}
%%%%%%%%%%%%%%%%%%%%%%%%%%%%%%%%%%%%%%%%%%%%%%%%%%%%%%%%%%

We begin our analysis of the isentropic curves by discussing the
$J=0$ limit of a collection of independent spin-1 sites.
The partition function is,
\begin{eqnarray}
Z = 1 + 2 e^{-\beta D} \,\,,
\label{eq:Jeq0a}
\end{eqnarray}
from which we can derive the internal energy, free energy, and entropy,
\begin{eqnarray}
\langle E \rangle &=& Z^{-1} 2 D e^{-\beta D}
\nonumber \\
F &=& -\frac{1}{\beta} \, {\rm ln} \, Z
\nonumber \\
S &=& \beta (\langle E \rangle -F) \,\,.
\label{eq:Jeq0b}
\end{eqnarray}
Since only the combination $\beta D$ enters the
expressions in Eq.~\ref{eq:Jeq0b},
the isentropic condition, $S=S_0$ implies that
$\beta D = c$, where the constant $c$ is obtained by
solving the transcendental equation,
\begin{eqnarray}
S_0 = \frac{2c \, e^{-c}}{1+2e^{-c}} + \ln(1+2e^{-c}) \,\,.
\end{eqnarray}

We immediately
see that the isentropic curves are straight lines
$D = c T$.
Note that for values of $S_0$ greater than $\ln2 \sim .693$, there
are two solutions for $c = D/T$, one positive and one negative,
representing positive and negative values of $D$.  In this
single site limit,
values of $D$ less than zero will suppress all vacancies, preventing entropy
values less than $\ln2$.

Despite the simple nature of this calculation, it allows
us to make some immediate statements about adiabatic heating and cooling.
We see that in traveling along an isentrope of
increasing $D$ from $D=-\infty$, the temperature $T$
decreases, since one moves along a line with $c<0$.
Ultimately, one reaches $D = 0$ where the sign of $c$ changes
to positive, and further movement along the isentrope results
in an increase in temperature.

The interesting question is, of course, how the topology
of the isentropes of the noninteracting systems is altered
by interactions and, in particular by the strong collective effects
which occur near the magnetic phase boundary.

%%%%%%%%%%%%%%%%%%%%%%%%%%%%%%%%%%%%%%%%%%%%%%%%%%%%%%%%%%
\subsection*{Mean Field Theory}
%%%%%%%%%%%%%%%%%%%%%%%%%%%%%%%%%%%%%%%%%%%%%%%%%%%%%%%%%%

\begin{figure}[t]
\includegraphics[width=8cm]{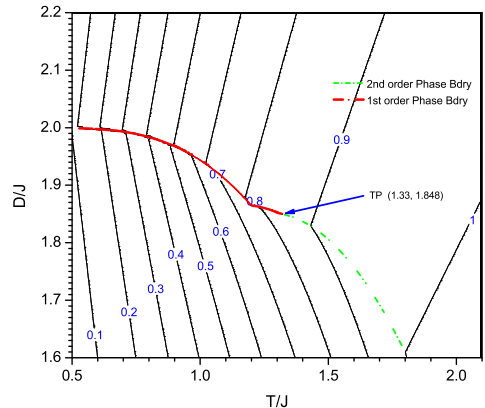}
\caption{(color online).
Mean-Field Theory calculation of the
isentropes of the Blume-Capel model.
The temperature $T$ decreases as $D$ increases
along the isentropes in the ordered phase,
and then rises in the paramagnetic phase.  $S$ jumps
at the phase boundary below the tricritical point, and
is continuous above.
}
\label{fig:BC_MFT_entropy}
\end{figure}

Having discussed the $J=0$ limit, we now
address the isentropic curves of the BCM treated within MFT.
Solving the self-consistency Eq~\ref{eq:selfconsist},
we show the resulting isentropes in Fig.~\ref{fig:BC_MFT_entropy}
along with the
mean-field phase boundary (PB). The PB consists of a first order
line (shown in full red) and second order boundary (dashed green),
which meet at the
tricritical point (TP):
$(T_{\rm tp}, D_{\rm tp}) = (\frac{4}{3},\frac{8}{3} \ln2)$. The
TP and  second order PB can be obtained analytically
by expanding the free energy
in powers of the magnetization $m$ and computing the Landau coefficient
for the $m^2$ term, thus fixing the transition line.
(See Ref.~[\onlinecite{blume66}].)
%% We estimated the first order PB by fixing the points at which the
%% order parameter $m$ vanished, which is a very sharp transition within MFT
%% with no local
%% fluctuations(all order is long range, by definition).

At the first order phase boundary, the entropy shows a characteristic
``jump," or discontinuity, as the isentropes take a sharp jog to the
left (smaller $T$) in passing through the boundary
(Fig.~\ref{fig:BC_MFT_entropy}).  The entropy decreases in this
traversal of the PB to higher $D$ values at fixed $T$ due to the large
reduction in the spin density $\rho$.  Below the PB in the ferromagnetic
phase (F), spin-0 ``vacancies" are sparse, resulting in an effective
(Ising-like) two spin state region and lower entropy.  Above the PB, all
three spin states are present with higher entropy.

In contrast, at the second order boundary (dashed green line in
Fig.~\ref{fig:BC_MFT_entropy}), the isentropes are continuous, but exhibit a
change in slope at the PB.  As we will see shortly, the precise details
of this behavior are
peculiar to the MFT and do not carry over to the exact (MC) solutions
which incorporate short range correlations and fluctuations.
Nevertheless the general topology of the isentropes in MFT agree
with those of MC.

%% For both first and second order transitions, MFT isentropes exhibit cooling,
%% (lower temperatures) with increasing D values below the PB, and heating(higher
%% temperatures) with increasing D above the PB.

%%%%%%%%%%%%%%%%%%%%%%%%%%%%%%%%%%%%%%%%%%%%%%%%%%%%%%%%%%
\subsection*{Monte Carlo- Wang-Landau}
%%%%%%%%%%%%%%%%%%%%%%%%%%%%%%%%%%%%%%%%%%%%%%%%%%%%%%%%%%

The phase diagram of the BCM was determined
by first using the WL density of states to compute
thermodynamic variables, the free energy $f$, entropy $s$, specific
heat $C(T)$, and magnetic susceptibility $\chi(T)$, for a fine grid
of points in the ($D$,$T$)-plane. We then examined specific
conditions to locate the phase transition points. For 1st order
transitions, we looked for jumps in the entropy coinciding with
a vanishing magnetic order parameter ($m$). For 2nd order transitions, we
located peaks in specific heat and  magnetic susceptibility curves
plotted as a function of temperature for fixed $D$.

\begin{figure}[t]
\includegraphics[width=8cm]{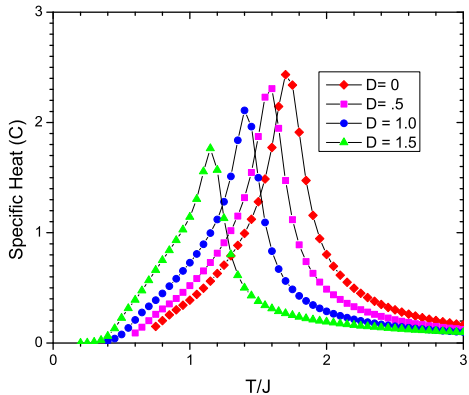}
\caption{(color online).
Wang-Landau MC results for the
specific Heat $C(T)$ for different values of the vacancy chemical
potential $D=0.0, 0.5, 1.0, 1.5$ in the second order phase transition region.
The peak positions give one estimate of the phase boundary.
}
\label{fig:BC_MC_SpecHeat}
\end{figure}

Representative results are shown in Fig.~\ref{fig:BC_MC_SpecHeat}.
Here $C(T)$ is obtained for fixed impurity chemical potentials
$D=0.0, 0.5, 1.0,$ and $1.5$.
Peaks are observed at temperatures
which are associated with the large fluctuations at the
second order phase transition.
Finite size scaling, and other sophisticated
techniques can be exploited to locate $T_c$ very precisely.
For example,
as $D$ is increased further, it is known that a tricritical
point exists in the BCM at $(T_{\rm tp}, D_{\rm tp})=(.610,1.965)$.
This change in behavior can be seen numerically by monitoring,
the appearance of hysteresis loops when MC simulations are
done sweeping $D$ at fixed $T< T_{\rm tp}$.

\begin{figure}[t]
\includegraphics[width=8cm]{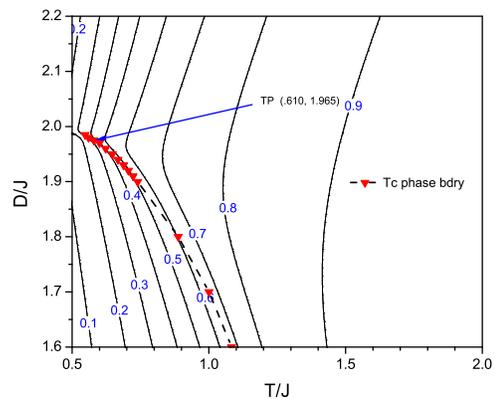}
\caption{(color online).
Wang-Landau MC results for the isentropes of the Blume-Capel model.
As for the MFT calculation, the isentropes exhibit
adiabatic cooling in the ferromagnetic phase, and heating in the
paramagnetic phase.  The entropy curves become discontinuous
at the tricritical point.
}
\label{fig:BC_MC_entropy_bw}
\end{figure}

Plots like that of Fig.~\ref{fig:BC_MC_SpecHeat} for different values of
$D$, as well as sweeps in $D$ at fixed $T$, were used to locate the phase
transition line in the $T-D$ plane.  This is shown as the
black dashed line in Fig.~\ref{fig:BC_MC_entropy_bw}.  The phase boundary
thus obtained is in good agreement (less than a percent difference)
with published results.  The new aspect of
Fig.~\ref{fig:BC_MC_entropy_bw}
is the inclusion of the isentropic curves.

The gross linear structure of the isentropes for large
$T/J$ and $D/J$ is well explained
by the $J=0$ (single site) analysis earlier in this section.
However, at intermediate $T/J$ and $D/J$,
the isentropes are rounded, especially in the vicinity of
the phase boundary.  Indeed, the boundary between adiabatic
heating and cooling does not precisely follow the PB but instead
occurs along a separate trajectory of somewhat larger impurity
chemical potential.

As discussed in [\onlinecite{mahmud10}],
the key qualitative feature of the isentropes
is that they move out to higher $T$ as they leave either side of the
phase boundary (into the paramagnetic or ferromagnetic phases).
The simple picture of this result is that the boundary
represents a line of a high degree of competition between different
phases, and hence a high entropy $S$.  In order for $S$ to remain
constant as we move away from the boundary, the temperature must
increase.  If an experiment were performed in which $D$ were ramped,
the lattice would cool as the phase boundary is approached
from the ferromagnetic side, and heat as one moves away into
the paramagnet.

\begin{figure}[t]
\includegraphics[width=8cm]{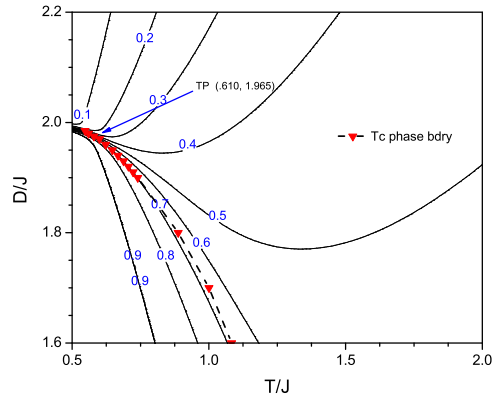}
\caption{(color online).
Wang-Landau MC results for the
spin density contours of the Blume-Capel model.
The vacancy density increases as the temperature $T$ or chemical
potential $D$ rise.  Ferromagnetic order is lost when roughly
one third of the sites are vacant.
}
\label{fig:BC_MC_spindensity}
\end{figure}

Comparing the isentrope contours, Fig.~\ref{fig:BC_MC_entropy_bw}, with the
those for spin density as shown in Fig.~\ref{fig:BC_MC_spindensity}, we see
that as we traverse the first order phase boundary, the spin density
changes more rapidly as the temperature $T$ approaches zero.  In fact,
this is consistent with Clausius-Clapeyron equation
which relates the slope of the phase boundary
with the change in entropy and spin density by,
\begin{eqnarray}
\frac{dD}{dT} = - \frac{s_{\rm fm} -s_{\rm pm}}{\rho_{\rm fm} -\rho_{\rm pm}}
\end{eqnarray}
where $s_{\rm fm}$ and $s_{\rm pm}$ stand for entropy in the ferromagnetic and
paramagnetic phases respectively.
We have verified that our results satisfy this
condition quantitatively.

%%%%%%%%%%%%%%%%%%%%%%%%%%%%%%%%%%%%%%%%%%%%%%%%%%%%%%%%%%
%%%%%%%%%%%%%%%%%%%%%%%%%%%%%%%%%%%%%%%%%%%%%%%%%%%%%%%%%%
\section*{IV.  Isentropic Curves of the Hubbard Hamiltonian}
%%%%%%%%%%%%%%%%%%%%%%%%%%%%%%%%%%%%%%%%%%%%%%%%%%%%%%%%%%
%%%%%%%%%%%%%%%%%%%%%%%%%%%%%%%%%%%%%%%%%%%%%%%%%%%%%%%%%%

\begin{figure}[t]
\epsfig{figure=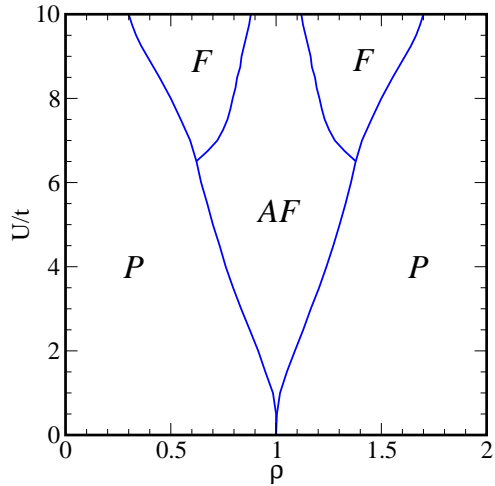,width=8cm,angle=-90,clip}
\caption{(color online).
Ground state MFT phase diagram of the two dimensional
square lattice Hubbard model.  Antiferromagnetism is
favored at and near half-filling, and extends all the way to
$U=0$ as a consequence of the divergence of the $\rho=1$
density of states.  Ferromagnetic regions are present at stronger
coupling.
}
\label{fig:Hirsch1}
\end{figure}

We now turn to the isentropic curves of the Hubbard Hamiltonian.
The ground state MFT phase diagram is shown in Fig.~\ref{fig:Hirsch1},
and consists of paramagnetic $P$ regions adjacent
to empty and fully occupied fillings ($\rho=0, 2$).
In the center, closer to half-filling, magnetic phases arise,
with antiferromagnetism {\cal AF} predominating immediately adjacent to $\rho=1$ and
ferromagnetism {\cal F} a bit farther away, at sufficiently large $U$.
The phase diagram is symmetric about $\rho=1$, as a consequence
of the particle-hole symmetry of the Hubbard model on a bipartite lattice
with near-neighbor hopping.

\begin{figure}[t]
\hskip1.0in
%% \centerline{\epsfig{figure=isenn1.0.ps,width=8cm,angle=-90,clip}}
\epsfig{figure=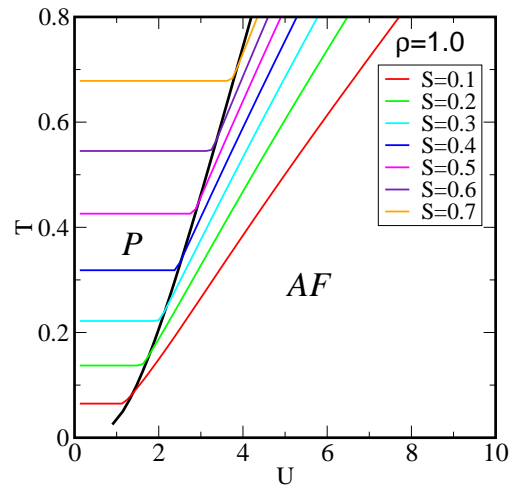,width=8cm,angle=-90,clip}
\caption{(color online).
Isentropic curves of the half-filled square lattice Hubbard model.
$S$ is independent of $U$ in the paramagnetic phase.  The
isentropic curves are continuous and bend upwards upon entering the ordered
antiferromagnetic region.  The entire $T=0$ axis is AF.
(See text and Fig.~\ref{fig:Hirsch1}.)
}
\label{fig:isenn1.0}
\end{figure}

\begin{figure}[t]
%% \centerline{\epsfig{figure=isenn0.8.ps,width=7cm,clip}}
\hskip1.0in
\epsfig{figure=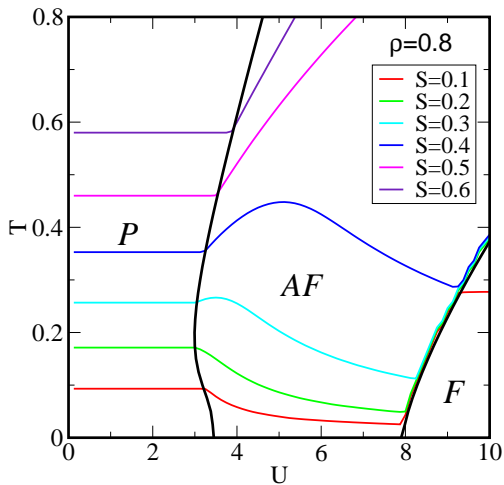,width=8cm,angle=-90,clip}
\caption{(color online).
Isentropic curves of the square lattice Hubbard model at
density $\rho=0.8$.
As at $\rho=1.0$ (Fig.~\ref{fig:isenn1.0}),
$S$ is independent of $U$ in the paramagnetic phase.  The
isentropic curves are continuous upwards upon entering the ordered
antiferromagnetic region, trending upwards for large $S$ and downwards
for small $S$.
The isentropic curves are discontinuous at the
{\cal AF} $\rightarrow$ {\cal F} transition.
As $U$ is increased at $T=0$, there is first a {\cal P} $\rightarrow$
{\cal AF}
transition, followed by a {\cal AF} $\rightarrow$ {\cal F} transition.
(See text and Fig.~\ref{fig:Hirsch1}.)
}
\label{fig:isenn0.8}
\end{figure}

Figure \ref{fig:isenn1.0}
shows the phase boundary in  the $T/t-U/t$ plane
at half-filling $\rho=1$.  As discussed earlier, the
associated isentropes are parallel to the $U/t$ axis in the
paramagnetic phase.  They then bend to higher $T/t$ as $U/t$
increases in the antiferromagnet.  This adiabatic heating is
explained by the same reasoning as for the classical BCM case:
The phase boundary represents a location of particularly
high entropy, so that for $S$ to remain fixed as one leaves its vicinity
the temperature must increase.  The gross morphology for the
isentropes found in
Fig.~\ref{fig:isenn1.0}
agrees well with exact QMC calculations\cite{paiva10}
which can be performed with no sign problem in this
half-filled case.  The QMC method, which is exact, of course captures the
fact that there is no finite temperature phase transition in the
square lattice FHM.  The role of the phase boundary there is played
by the temperature scale of the mean field charge gap,
which is determined by the plateau in $\rho$ versus $\mu$.

Figure \ref{fig:isenn0.8} shows the analogous phase boundaries and
isentropes for $\rho=0.8$.  Here the structure is much
richer since there are three possible phases for this density,
with the paramagnet first giving way to antiferromagnetic order
as $U/t$ is increased, followed by a second transition
to ferromagnetism.  In this case the isentropes can bend either
to higher or to lower $T$ as the AF boundary is left with
increasing $U$.  The decrease in $T$ occurs at lower $T/t$
where the {\cal F} boundary is more proximate to the {\cal AF}
one.  The isentropes move to higher $T$ as $U/t$ decreases from
the {\cal F} phase boundary.
The other interesting feature of
Fig.~\ref{fig:isenn0.8},
not present in
Fig.~\ref{fig:isenn1.0},
is the discontinuity in the isentropes at the {\cal AF}-{\cal F}
boundary.  This occurs here, similar to the situation
for temperatures below the tricritical point in the
BCM model, because of the first-order nature of the transition.
As with the BCM,
we have verified that the entropy jump along the
first order boundary satisfies the Clausius-Clapeyron relation,
which here takes the form,
%\color{blue}
\begin{eqnarray}
\frac{dU}{dT} = \frac{s_{\rm fm} -s_{\rm afm}}
{{\cal D}_{\rm fm} -{\cal D}_{\rm afm}}
\,\,.
\end{eqnarray}
Here $\frac{dU}{dT}$ is the slope of the AF-F phase boundary,
$s_{\rm fm}, s_{\rm afm}$ are the associated entropies,
and
${{\cal D}_{\rm fm}, {\cal D}_{\rm afm}}$
are the double occupations in the two phases.
%\color{black}

\begin{figure}[t]
\epsfig{figure=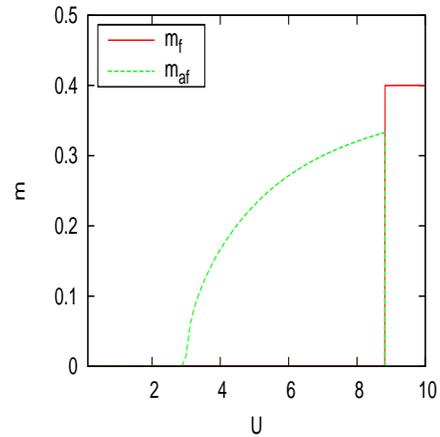,width=6cm,height=6cm,clip}
\caption{(color online).
The antiferro- and ferro-magnetic order parameters are shown for
$\rho=0.8$ and $T=0.2$.  As $U$ increases from zero there
is a continuous transition to an AF phase followed
by a discontinuous transition to a F phase.
}
\label{fig:n0.4T0.2mfmaf}
\end{figure}

The behavior of the ferromagnetic, $m_{\rm f}$, and antiferromagnetic,
$m_{\rm af}$, order parameters for $\rho=0.8$ and $T=0.2$ is shown in
Fig.~\ref{fig:n0.4T0.2mfmaf}.  Both change discontinuously through
the first order AF-F transition at $U \approx 8.8t$.  However at the second
order P-AF
transition at $U \approx 3t$, $\, m_{\rm af}$ increases continuously
from zero with the MFT exponent $\beta=\frac{1}{2}$.

An interesting feature of the {\cal P} $\rightarrow$
{\cal AF} phase boundary at $\rho=0.8$ is that the critical $U$
initially {\it decreases} as the temperature $T$ {\it increases}.
Put another way, as $T$ is lowered for $U \approx 3.2$ there
is a
{\cal P} $\rightarrow$ {\cal AF}
ordering transition, but then there is
a re-entrance to the {\cal P} phase as $T$
is reduced further.
We have verified that this phenomenon is obtained
also within an independent random phase approximation
calculation, in which the magnetic susceptibility is given by,
\begin{eqnarray}
\chi({\bf q},T) &=& \frac{\chi_0({\bf q},T)}{1-U\chi_0({\bf q},T)}
\nonumber \\
\chi_0({\bf q},T) &=& \sum_{\bf k}
\frac{ f(\epsilon_{{\bf k}+{\bf q}}) -f(\epsilon_{{\bf k}}) }
{ \epsilon_{{\bf k}} - \epsilon_{{\bf k}+{\bf q}}  } \,\, .
\label{eq:RPA}
\end{eqnarray}
The critical temperature is then determined from Eq.~\ref{eq:RPA}
via the Stoner criterion $1 - U \chi_0({\bf q},T) = 0$,
and exhibits the same re-entrant phenomenon as the MFT
calculation.

\begin{figure}[t]
\hskip1.0in
\epsfig{figure=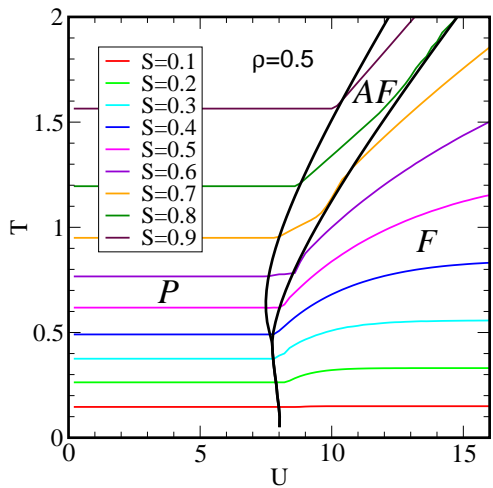,width=8cm,angle=-90,clip}
\caption{(color online).
Phase diagram and isentropes for quarter filling.
For $T > 0.5 t$,
an antiferromagnetic region appears between the
paramagnetic region at weak coupling and ferromagnetic region
at strong coupling.
}
\label{fig:isenn0.5}
\end{figure}

Finally, we show in Fig.~\ref{fig:isenn0.5} the phase
diagram and isentropes for quarter filling.
At zero temperature (see
Fig.~\ref{fig:Hirsch1}) there is a paramagnetic to
ferromagnetic transition as $U$ increases.
However, at higher $T$ an intermediate antiferromagnetic phase
intervenes.  The general trend is towards adiabatic
heating as $U$ rises.

%%%%%%%%%%%%%%%%%%%%%%%%%%%%%%%%%%%%%%%%%%%%%%%%%%%%%%%%%%
%%%%%%%%%%%%%%%%%%%%%%%%%%%%%%%%%%%%%%%%%%%%%%%%%%%%%%%%%%
\section*{V.  Conclusions}
%%%%%%%%%%%%%%%%%%%%%%%%%%%%%%%%%%%%%%%%%%%%%%%%%%%%%%%%%%
%%%%%%%%%%%%%%%%%%%%%%%%%%%%%%%%%%%%%%%%%%%%%%%%%%%%%%%%%%

In this paper, the isentropic curves of a classical system
of magnetism, the Blume-Capel model, and of the quantum fermion
Hubbard Hamiltonian, have been determined.

In the case of the BCM, Mean Field Theory and Monte Carlo
(Wang-Landau) calculations give a qualitatively similar pictures
in which adiabatic heating is observed as one moves away from the
phase boundary,
although the precise, quantitative location of the transition is,
of course, different in the two methods.
The behavior of the isentropes is made somewhat more
complex by the presence of a tricritical point on
the BCM phase boundary, so that there is a region
at low $T$ and large vacancy fraction where the
curves are discontinuous.

For the FHM, we have presented only MFT results, since the sign problem
in general prevents Quantum Monte Carlo simulations at low temperatures.
The topology of the isentropes, like the phase diagram itself, is
relatively simple at half-filling, consisting of lines parallel to
the $U$ axis in the paramagnetic phase, and then trending upwards as the
antiferromagnetic boundary is crossed.
For this half-filled
case, QMC is possible, and we have compared our MFT results
to those calculations.
The isentropes are considerably more elaborate in the doped case, where
MFT exhibits competing transitions to ferro- and antiferro-magnetic
orders.  In these cases the dependence of the temperature along the
isentropes is non-monotonic.

It is well known that MFT predicts some qualitatively incorrect
features of the FHM phase diagram, including, for example,
the existence of a finite temperature Ne\'el transition.
Nevertheless, the rough morphology of the isentropes
within MFT and QMC are similar, with the role of the
MFT phase boundary played by the temperature of the charge gap, which is
clearly seen to open at nonzero temperatures.
This provides some assurance that
the MFT results will remain qualitatively accurate in the doped
case, even in the absence of exact
QMC results at low temperatures there.

%\color{blue}
Optical lattice experiments for
bosons\cite{spielman0708,jimenez10,gemelke09,trotzky09} and
fermions\cite{jordens08,schneider08} have reached temperatures
($T \simeq t$ in the bosonic case) that provide strong evidence of the
Mott transition.  A reduction in the experimentally accessible entropy
per fermionic atom, $s \simeq {\rm ln}2$, by a factor of
2-3 will allow the observation of local spin correlations.\cite{paiva10}
%% Hence the development of additional cooling methods is central to the field.
One possibility raised by the results of Fig.~\ref{fig:isenn0.8} is that
adiabatic cooling can occur in the proximity of two competing types of
order.  However, Fig.~\ref{fig:isenn0.8} also suggests that this cooling
only occurs when $T$ is already sufficiently low.
%\color{black}

We acknowledge financial support from ARO Award W911NF0710576
with funds from the DARPA OLE program.

\end{document}